\documentclass[12pt]{article}

\usepackage{color}
\usepackage[hidelinks,colorlinks,allcolors=blue]{hyperref}
\usepackage{amssymb}
\usepackage{amsmath}
\usepackage{array}
\usepackage{graphicx}
\usepackage{enumitem}

\usepackage{point}
\usepackage{largearray}

\usepackage{dsfont}
\usepackage{theorem}

\newtheorem{Theorem}{Theorem}
\newtheorem{Lemma}[Theorem]{Lemma}
\numberwithin{equation}{section}
\numberwithin{Theorem}{section}

\def\indent{\hskip20pt}
\def\qed{\hfill$\square$}

\begin{document}

\pagestyle{empty}

\hbox{}
\hfil{\bf\LARGE Analysis of a simple equation for the
\par
\vskip8pt
\hfil ground state energy of the Bose gas
}
\vskip10pt

\hfil{\bf Eric A. Carlen}\par
\hfil{\it Department of Mathematics, Rutgers University}\par
\hfil{\tt\href{mailto:carlen@rutgers.edu}{carlen@rutgers.edu}}\par
\vskip5pt

\hfil{\bf Ian Jauslin}\par
\hfil{\it Department of Physics, Princeton University}\par
\hfil{\tt\href{mailto:ijauslin@princeton.edu}{ijauslin@princeton.edu}}\par
\vskip5pt

\hfil{\bf Elliott H. Lieb}\par
\hfil{\it Departments of Mathematics and Physics, Princeton University}\par
\hfil{\tt\href{mailto:lieb@princeton.edu}{lieb@princeton.edu}}\par

\begin{abstract}
In 1963 a partial differential equation with a convolution non-linearity was introduced in connection with a quantum mechanical many-body problem, namely the gas of bosonic particles.
This equation is mathematically interesting for several reasons.
(1) Although the equation was expected to be valid only for small values of the parameters, further investigation showed that predictions based on the equation agree well over the {\it entire range} of parameters with what is expected to be true for the solution of the true many-body problem.
(2) The novel nonlinearity is easy to state but seems to have almost no literature up to now.
(3) The earlier work did not prove existence and uniqueness of a solution, which we provide here along with properties of the solution such as decay at infinity. 
\end{abstract}

{\small
\tableofcontents
}
 
\vfill
\hfil{\scriptsize\copyright\ by the authors. This paper may be reproduced in its entirety for non-commercial purposes.}
\eject

\setcounter{page}1
\pagestyle{plain}

\section{Introduction}
\indent This paper is devoted to the study of an integro-differential equation introduced in\-~\cite{Li63} in connection with the study of the Bose gas, a many body problem in quantum mechanics.
The equation is
\begin{equation}
 \boxed{\phantom{\int}(-\Delta+4e+\mathcal V(x))u(x)=\mathcal V(x)+2e\rho(u\ast u)(x)\ ,\phantom{\int} }
 \label{simpleq}
\end{equation}
with $x\in\mathbb R^d$, and $*$ denoting convolution: $u\ast u(x):=\int u(x-y)u(y)\ dy$.
Here, $\mathcal V$ is a given function, (called the {\it potential}), in $L^1 (\mathbb{R}^d ) \cap L^p(\mathbb{R}^d )$,
with $p >d/2$ for $d\geqslant 2$ and $p>1 $for $d=1$. We assume $\mathcal{V}$ to be non-negative. (This corresponds to a repulsive interaction between the particles in the underlying quantum system). The two parameters
$e$ and $\rho$ are non-negative numbers, and they are related by a constraint, namely
\begin{equation}
 e=\frac\rho2\int (1-u(x))\mathcal V(x)\ dx\ .
 \label{energy}
\end{equation}

\indent We are interested in solutions of\-~(\ref{simpleq}) that satisfy the constraint\-~(\ref{energy}), or, in other words, solutions of the system\-~(\ref{simpleq}) and\-~(\ref{energy}). We are particularly interested in the case $d=3$, though other dimensions are also of interest. 
As explained in \-~\cite{Li63}, the parameter $\rho$ corresponds to the particle density $\frac NV$ of the underlying Bose gas in the large volume and large particle number limit, and $e=\frac EN$ stands for the energy per particle. 
\bigskip

\indent One would like to fix a value $\rho$ for the density, and then one expects, on the basis of the arguments in \-~\cite{Li63}, that there will be a unique value of $e = e(\rho)$ such that there is a solution of\-~(\ref{simpleq}) and\-~(\ref{energy}) with $u$ taking values in $[0,1]$.
This value of $e$ is then the energy per particle of the Bose gas in its ground state.

\indent The problem of determining this ground state energy per particle, as a function of the density, has attracted the attention of a great many researchers since the pioneering work of Lenz in 1929\-~\cite{Le29}.
In that paper and subsequent work\-~\cite{Bo47,LHY57}, an asymptotic expansion of $e(\rho)$, for $d=3$ and small $\rho$ was obtained:
\begin{equation}
 e=2\pi\rho a\left(1+\frac{128}{15\sqrt\pi}\sqrt{\rho a^3}+o(\sqrt\rho)\right)\ \qquad \rho \to 0
 \label{lhy}
\end{equation}
where $a$, called the {\it scattering length}, is a property of the pair interaction potential $\mathcal V(x)$, and is defined in\-~(\ref{scattering})-(\ref{ephi}) below.
Here, we set both the mass $m$ of the particle and Planck's constant $\hbar$ to 1. This early work was not mathematically rigorous, and it was not until
 1998\-~\cite{LY98} that the validity of the first term $2\pi\rho a$ was proved, and not until 2019\-~\cite{FS19} that the validity of second term was also proved, utilizing upper bounds proved earlier in\-~\cite{Dy57,YY09}.

\indent This timeline gives some idea of the complexity of the problem of directly studying the Bose gas ground state as a many body problem. The complexity makes it very attractive to try to show that the system\-~(\ref{simpleq}) and 
 (\ref{energy})
provides a useful and illuminating route to the computation of the properties of the ground state for a Bose gas. Interest is piqued further by the fact that numerical studies show that the function $e(\rho)$ computed using the system\-~(\ref{simpleq}) and\-~(\ref{energy}) is surprisingly accurate for {\em all } densities, not only low densities, as we discuss later in this paper. Until now, however, there has been no mathematically rigorous study of this system, and even the most basic questions concerning existence and uniqueness of solutions had remained open. 
\bigskip

\indent In this paper, we settle some of these basic questions and raise others. It may at first appear surprising that the equation\-~(\ref{simpleq}) poses any serious mathematical challenges. After all, if one replaced the convolution nonlinearity $u*u$ in\-~(\ref{simpleq}) by a power non-linearity, say $u^2$, one would have a familiar sort of local elliptic equation:
\begin{equation}
 (-\Delta+4e+\mathcal V(x))u(x)=\mathcal V(x)+2e\rho u^2(x)
 .
 \label{simpleq2}
\end{equation}
However, the convolution nonlinearity in\-~(\ref{simpleq}) makes it non-local, and very different from\-~(\ref{simpleq2}).
\bigskip

\indent As explained in\-~\cite{Li63} the solutions of physical interest are integrable and {\em must} satisfy  $u(x) \leqslant 1$ for all $x$. Our first result is  that for  integrable solutions of  the  system\-~(\ref{simpleq})-(\ref{energy}), the upper bound $u \leqslant 1$ implies the lower bound $u \geqslant 0$:

\begin{Theorem}[Positivity]\label{positivity}   Suppose that $\mathcal{V}$ is  non-negative and integrable and that $u$ is an integrable solution of 
(\ref{simpleq})-(\ref{energy}) such that  $u(x) \leqslant 1$ for all $x$. Then $u(x) \geqslant 0$ for all $x$, and all such solutions have fairly slow decay at infinity in that they satisfy
\begin{equation}\label{slow}
\int |x|u(x)d x = \infty \ .
\end{equation}
Thus, any physical solutions of (\ref{simpleq})-(\ref{energy}) must necessarily satisfy the {\em pair} of inequalities
\begin{equation}
  0  \leqslant u(x) \leqslant 1 \quad{\rm for \ all}\ x \ .
  \label{con1}
\end{equation}
\end{Theorem}

\bigskip
This {\em a-priori} result, which we prove before we take up existence and uniqueness, turns on results \cite{CJLL20}   obtained in collaboration with Michael Loss on the convolution inequality $f \geqslant f\ast f$ in $L^1(\mathbb R^d)$.
While $u(x)\leqslant 1$ is a physical requirement, $u(x)\geqslant0$ is not, see section\-~\ref{sec:bosegas} for details.
\bigskip

The converse of Theorem\-~\ref{positivity} also holds, as stated in the following theorem.

\begin{Theorem}\label{theorem:leq1}
  Let $\mathcal V\in L^1(\mathbb{R}^d)\cap L^p(\mathbb{R}^d)$, $p>\max\{\frac d2,1\}$, be non-negative.
  If $u$ is an integrable solution of 
(\ref{simpleq})-(\ref{energy}) such that  $u(x) \geqslant 0$ for all $x$, then $u(x) \leqslant 1$ for all $x$.
\end{Theorem}

{\bf Remark}: We have thus proved that $u\geqslant0$ if and only if $u\leqslant1$.
This, in principle, leaves the door open to solutions that are sometimes $>1$ and sometimes $<0$, though we do not believe such solutions exist.
\bigskip

\indent Before stating our main theorems, we make a few observations. 
\medskip

\point The system\-~(\ref{simpleq})-(\ref{energy}) is actually equivalent to\-~(\ref{simpleq}) and the constraint
\begin{equation}
 \int u(x)\ dx=\frac1\rho.
 \label{con2}
\end{equation}
To prove this, consider the operator
\begin{equation}
 G_e := [-\Delta + 4e]^{-1}
 \label{Ge}
\end{equation}
which is given by
\begin{equation}
 G_e f = Y_{4e}*f
 \label{yukawa}
\end{equation}
where $Y_{4e}$ is the {\em Yukawa potential}\-~\cite[section 6.23]{LL01}, which is non-negative and $\int Y_{4e} dx = (4e)^{-1}$.
When $d=3$,
\begin{equation}
 Y_{4e}(x)=\frac{e^{-2\sqrt e|x|}}{4\pi|x|}
 .
\end{equation}
Equation\-~(\ref{simpleq}) can be rewritten as
\begin{equation}\label{simpleq3}
u(x) = Y_{4e}*(\mathcal V (1- u(x))) +2e\rho Y_{4e}*u*u \ .
\end{equation}
Since $u$ and $\mathcal V$ are assumed to be integrable, and $u(x)$ is assumed to satisfy\-~(\ref{con1}), all terms in\-~(\ref{simpleq3}) are integrable, and integrating yields 
\begin{equation}
\int u(x)\ dx = \frac{1}{4e} \int \mathcal V(x)(1- u(x)) dx +\frac{\rho}{2} \left(\int u(x)\ dx\right)^2\ .
\end{equation}
Thus, for integrable solutions $u$ of\-~(\ref{simpleq}) satisfying\-~(\ref{con1}), the constraint\-~(\ref{energy}) is equivalent to\-~(\ref{con2}).
\bigskip

\point
There is another useful way to write the system \-~(\ref{simpleq})-(\ref{energy}). The damped heat semigroup $e^{-t(-\Delta + 4e)}$ is a strongly continuous contraction semigroup on $L^p(\mathbb{R}^d)$, and the domain of its generator is 
$\mathcal{D}(-\Delta + 4e)= W^{2,p}(\mathbb{R}^d)$. 
By the Sobolev embedding theorem\-~\cite[Theorem 10.2]{LL01}, since $p>d/2$, all functions $f\in \mathcal{D}(-\Delta + 4e)$ are continuous and vanish at infinity. Since $\mathcal{V} \geqslant 0$, $e^{-t\mathcal{V}}$ is also a
strongly continuous contraction semigroup on $L^p(\mathbb{R}^d)$, and since $\mathcal{V} \in L^p(\mathbb R^d)$, the domain of its generator, $\mathcal{D}(\mathcal{V})$, contains all bounded functions, and in particular
$W^{2,p}(\mathbb{R}^d)$. Writing $\mathcal{V}$ as the sum of a piece with a small norm in $L^p(\mathbb R^d)$ and another piece that is bounded, it is easy to see that there are numbers $a,b>0$ with $a< 1/2$ such that for all $f\in 
W^{2,p}(\mathbb{R}^d)$, 
\begin{equation}
 \|\mathcal{V}f\|_p \leqslant a\|(-\Delta + 4e)f\|_p + b\|f\|_p\ .
\end{equation}
Then by the Banach space version
of the Kato-Rellich theorem, \cite[p.~244]{RS75b}, the operator $-\Delta + 4e + \mathcal V(x)$ maps 
$W^{2,p}(\mathbb{R}^d)$ invertibly onto $L^p(\mathbb R^d)$. Define $K_e$ to be the inverse operator
\begin{equation}
 K_e := [-\Delta + 4e + \mathcal V(x)]^{-1}\ .
 \label{Kedef}
\end{equation}
By the Trotter product formula, the operator $K_e$ has a positive kernel that we denote by $K_e(x,y)$; in particular, $K_e$ preserves positivity.
By the resolvent identity
\begin{equation}
 K_e=G_e-G_e\mathcal VK_e
 \label{resolvent}
\end{equation}
we conclude that
\begin{equation}
 0\leqslant K_e(x,y)\leqslant G_e(x,y)
 \label{ineqKe}
\end{equation}
for all $x,y$. Thus, the operator $K_e$ extends to a bounded operator on $L^1(\mathbb R^d)$ and all terms in the equation 
\begin{equation}\label{simpleq4}
u(x) = K_e \mathcal V(x) +2e\rho K_eu*u(x) \ .
\end{equation}
are well-defined whenever $u$ is integrable. Moreover, since
$\mathcal V \in L^p(\mathbb R^d)$, and since $u\ast u\in L^p(\mathbb R^d)$ when $u$ is integrable and satisfies\-~(\ref{con1}), every integrable solution $u$ of\-~(\ref{simpleq4}) that satisfies\-~(\ref{con1}) actually belongs to
$W^{2,p}(\mathbb{R}^d)$ and satisfies\-~(\ref{simpleq}).
\bigskip

\indent Several simple bounds follow almost immediately from this form of the equation.
First of all, since the last term on the right of\-~(\ref{simpleq4}) is non-negative, we have an {\em a-priori} lower bound on $u(x)$, namely
\begin{equation}
 u(x) \geqslant u_1(x) := K_e\mathcal V(x)\ .
 \label{u1def}
\end{equation}
Integrating both sides of\-~(\ref{u1def}), and using\-~(\ref{con2}) yields an upper bound on $\rho$ depending only on $e$, namely, $\rho \leqslant \left(\int K_e\mathcal V(x)\ dx\right)^{-1}$. 
By\-~(\ref{energy}) and\-~(\ref{u1def}),
\begin{equation}
\rho = {2e}\left(\int\mathcal{V}(1- u)(x) dx\right)^{-1} \geqslant {2e}\left(\int\mathcal{V}(1- K_e\mathcal V)(x) dx\right)^{-1} \ .
\end{equation}
Altogether,
\begin{equation}
{2e}\left(\int\mathcal{V}(1- K_e\mathcal V)(x) dx\right)^{-1} \leqslant \rho \leqslant \left(\int K_e\mathcal V(x)\ dx\right)^{-1} \ .
 \label{con4}
\end{equation}
In fact, the left side of\-~(\ref{con4}) is equal to one half the right side. To see this observe that $u_1 = K_d \mathcal{V}$ satisfies $(-\Delta +4e + \mathcal{V})u_1 = \mathcal{V}$, and hence $u_1 = G_e(\mathcal{V}(1-u_1))$.
Integrating both side yields $\int u_1 dx = \frac{1}{4e} \int \mathcal{V}(1-u_1) dx$. By\-~(\ref{u1def}), we obtain the following simpler (albeit less sharp) bounds:
\begin{equation}
{2e}\left(\int\mathcal{V}dx \right)^{-1} \leqslant \rho \leqslant{4e}\left(\int\mathcal{V}dx \right)^{-1} \ ,
 \label{con4B}
\end{equation}
or equivalently
\begin{equation}
\left( \frac{1}{4}\int\mathcal{V}dx \right)\rho \leqslant e \leqslant \left(\frac{1}{2}\int\mathcal{V}dx\right)\rho \ .
 \label{con4C}
\end{equation}
In particular, this shows that the system\-~(\ref{simpleq})-(\ref{energy}) does not have a solution for arbitrary values of $\rho$ and $e$: when either is small, a solution of the type we seek can only exist if the other is correspondingly small, as specified by\-~(\ref{con4B}) and\-~(\ref{con4C}).
In fact, as is stated in the following theorem, $\rho$ and $e$ are constrained to be related by a functional equation.
\bigskip

\begin{Theorem}[existence and uniqueness]\label{theorem:existence} Let $\mathcal V\in L^1(\mathbb{R}^d)\cap L^p(\mathbb{R}^d)$, $p>\max\{\frac d2,1\}$, be non-negative. 
Then there is a constructively defined continuous function $\rho(e)$ on $(0,\infty)$ such that 
 $\lim_{e\to 0}\rho(e) = 0$ and $\lim_{e\to \infty} \rho(e) = \infty$ and such that for any $e\geqslant 0$ and $\rho = \rho(e)$, 
 the system\-~(\ref{simpleq}) and\-~(\ref{energy}) has a unique integrable solution $u(x)$ satisfying $u(x)\leqslant 1$. Moreover, if $\rho \neq \rho(e)$, the system\-~(\ref{simpleq}) and\-~(\ref{energy}) has {\em no} integrable solution $u(x)$ satisfying\-~(\ref{con1}).
\end{Theorem}
\bigskip

{\bf Remark}:
\begin{itemize}[beginpenalty=10000]
 \item We do not assume here that the potential is radially symmetric.
 However, the uniqueness statement implies that $u$ is radially symmetric whenever $\mathcal V$ is radially symmetric.

 \item The function $\rho(e)$ is the {\em density function}, which specifies the density as a function of the energy. Thus, our system together with\-~(\ref{con1}) constrains the parameters $e$ and $\rho$ to be related by a strict functional relation $\rho = \rho(e)$. In most of the early literature on the Bose gas, $\rho$ is taken as the independent parameter, as suggested by\-~(\ref{lhy}): One puts $N$ particles in a box of volume $N/\rho$, and 
seeks to find the ground state energy per particle, $e$, as a function of $\rho$. Our theorem goes in the other direction, with $\rho$ specified as a function of $e$. We prove that $e\mapsto\rho(e)$ is continuous, and we conjecture that $\rho(e)$ is a strictly 
monotone increasing function. In that case, the functional relation could be inverted, and we would have a well-defined function $e(\rho)$.

 \item Since $ \lim_{e\to 0}\rho(e)=0$ and $\lim_{e\to \infty}\rho(e)=\infty$, the continuity of $e\to \rho(e) $ implies that for each $\rho \in (0,\infty)$ there is {\it at least one} $e$ such that $\rho(e) =\rho$.
\end{itemize}
\bigskip

\indent Having proved that the solution to the simple equation is unique, our second main result is an asymptotic expression for $e(\rho)$, both for low and for high density.
\bigskip

\begin{Theorem}[asymptotics of the energy for $d=3$]\label{theorem:asymptotics}
 Consider the case $d=3$.
 Let $\mathcal V$ be non-negative, integrable and square-integrable. Then, for each 
 $\rho>0$ there is at least one $e>0$ such that $\rho = \rho(e)$. For any such $\rho $ and $e$ we have the following bounds for low and high density (i.e., small and large $\rho$).
 For low density, 
 \begin{equation}
 \boxed{\ e=2\pi\rho a\left(1+\frac{128}{15\sqrt\pi}\sqrt{\rho a^3}+o(\sqrt\rho)\right)\phantom{F} }
 \label{3d}
 \end{equation}
 where $a$ is the {\it scattering length} of the potential, which is defined in~(\ref{sl}).
 For high density, in any dimension $d\geqslant 1$,
 \begin{equation}
 e=\frac\rho2\int\mathcal V(x)\ dx+o(\rho)
 .
 \label{largerho}
 \end{equation}
\end{Theorem}
\bigskip

{\bf Remarks}:
\begin{itemize}[beginpenalty=10000]
 \item For low densities in $d=3$, the energy $e$ predicted by the simple equation\-~(\ref{simpleq})-(\ref{energy}) is asymptotically equal to the ground state energy of the Bose gas\-~\cite{LHY57,YY09,FS19}.
 For high densities, when the potential has a non-negative Fourier transform, the asymptotic formula for the ground state energy of the Bose gas coincides with\-~(\ref{largerho})\-~\cite[appendix]{Li63}.
 Thus, the simple equation yields the same asymptotes for both low and high densities as the Bose gas does (at least when the potential has a non-negative Fourier transform, as in the example $\mathcal V(x)=e^{-|x|}$ discussed in section\-~\ref{subsec:numerics}).
\end{itemize}
\bigskip

\begin{Theorem}[decay of $u$ at infinity]\label{theorem:decay} In all dimensions, provided $\mathcal{V}$ is spherically symmetric with $\int |x|^2\mathcal{V} dx <\infty $ in addition to satisfying the hypotheses imposed in Theorem~\ref{theorem:existence}, all integrable solutions of (\ref{simpleq})-(\ref{energy}) with $u(x) \leqslant1$ for all $x$ satisfy
\begin{equation}\label{gendecay}
\int |x| u(x) dx = \infty \quad{\rm and}\quad \int |x|^r u(x) dx < \infty \quad{\rm for \ all} \ 0 < r < 1\ .
\end{equation}
Thus, if $u(x) \sim |x|^{-m}$ for some $m$, the only possibility is $m = d+1$. Under stronger assumptions on the potential, this is actually the case.
For $d=3$, if $\mathcal V$ is non-negative, square-integrable, spherically symmetric (that is, $\mathcal V(x)=\mathcal V(|x|)$), and, for $|x|>R$,
  \begin{equation}
    \mathcal V(|x|)\leqslant Ae^{-B|x|}
    \label{expdecay}
  \end{equation}
  for some $A,B>0$ then there exists $\alpha>0$ such that
  \begin{equation}
    u(x)\mathop\sim_{|x|\to\infty}\frac\alpha{|x|^4}   .
  \end{equation}
\end{Theorem}
\bigskip

{\bf Remarks}:
\begin{itemize}[beginpenalty=10000]
 \item This result is consistent with a prediction in\-~\cite{LHY57} that the truncated 2-point correlation function in the ground state of the Bose gas decays like $|x|^{-4}$.

 \item To prove this theorem, we will use analytical properties of the Fourier transform $\widehat{\mathcal V}$ of $\mathcal V$, which is why we assume that $\mathcal V$ decays exponentially at infinity.
 For potentials with slower decay, it seems that the decay of $u$ should still be $|x|^{-4}$, except if $\mathcal V$ itself decays slower than $|x|^{-4}$, in which case $u$ should decay like $\mathcal V$.

 \item It is presumably not too difficult to extend this result to cases with potentials that are not spherically symmetric.
\end{itemize}
\bigskip

{\bf Remark}:
The simple equation\-~(\ref{simpleq}) is actually an approximation of a richer equation for $u$\-~\cite{Li63}, which should more accurately depict the Bose gas, see\-~(\ref{fulleq}).
 Little is known about this richer equation.
\bigskip

\indent The paper is organized as follows.
We prove theorem\-~\ref{positivity} in section\-~\ref{sec:pos}, theorems\-~\ref{theorem:leq1} and\-~\ref{theorem:existence} in section\-~\ref{sec:existence}, theorem\-~\ref{theorem:asymptotics} in section\-~\ref{sec:asymptotics}, and theorem\-~\ref{theorem:decay} in section\-~\ref{sec:decay}.
In section\-~\ref{sec:bosegas}, we explain how the simple equation is related to the Bose gas, and present some numerical evidence that it is very good at predicting the ground state energy.
In section\-~\ref{sec:open} we discuss a few open problems and extensions.

\section{Proof of Theorem~\ref{positivity}}\label{sec:pos}

As explained in the introduction, the solutions of (\ref{simpleq})-(\ref{energy}) that are of physical interest are those that are integrable and satisfy $u(x) \leqslant 1$ for all $x$.  In this section we prove, making no  assumptions on the potential $\mathcal{V}$ other than its positivity and integrability, that all such solutions are non negative, and have slow decay so that $\int |x| u(x) dx = \infty$.   

Our starting point is the form of (\ref{simpleq}) given in (\ref{simpleq3}). For an integrable solution $u$, define
\begin{equation}\label{pos1}
f :=  2e\rho Y_{4e}*u\ .
\end{equation}
If (\ref{energy}) is satisfied,  then
\begin{equation}\label{pos2}
\int f dx = \frac 12\ .
\end{equation}
and (\ref{simpleq3}) can be written as
\begin{equation}\label{pos3}
u = Y_{4e}*(\mathcal V (1- u)) +  f\ast u\ .
\end{equation}

\begin{Lemma} Let $u(x)$ be an integrable solution of the system (\ref{simpleq})-(\ref{energy}) such that $u(x) \leqslant 1$ for all $x$. Let $f$ be defined in terms of $u$, $e$ and $\rho$ by (\ref{pos1}).  If $f(c) \geqslant 0$ for all $x$, then $u(x) \geqslant 0$ for all $x$.
\end{Lemma}

{\bf Proof} Since $Y_{4e}*(\mathcal V (1- u(x))) \geqslant 0$, it follows that 
\begin{equation}
  u_- \leqslant  (f*u)_- =  (f*u_+ - f*u_-)_-  \leqslant f*u_-\ .
\end{equation}
Integrating, we find
${\displaystyle \int u_- dx \leqslant \frac12 \int u_-dx}$,
and this implies that $u_- = 0$. \qed

{\bf Proof of Theorem~\ref{positivity}} Multiply (\ref{pos3}) through by $2e\rho$, and then convolve both sides with $Y_{4e}$. The result is  $f = 2e\rho Y_{4e}*(Y_{4e}*(\mathcal V (1- u)) +  f\ast f$, and since   $Y_{4e}*(Y_{4e}*(\mathcal V (1- u))) \geqslant 0$, $f$ is an integrable solution of 
\begin{equation}\label{pos5}
f(x) \geqslant f\ast f(x)
\end{equation}
for all $x$. It is proved in \cite{CJLL20} that all integrable solutions of (\ref{pos5}) are non-negative and have integral no greater than  $\tfrac12$, and that moreover, (\ref{pos2}) and (\ref{pos3}) together imply that 
\begin{equation}
  \int |x| f(x)\ dx = \infty\ .
\end{equation}
However, 
\begin{equation}
  \int |x| f(x)\ dx = 2e\rho\int |x| Y_{4e}\ast u(x)\ dx= 2e\rho\int (Y_{4e} \ast  |x|)  u(x)\ dx \ .
\end{equation}
Then since $\lim_{x\to \infty}\left(4e|x|^{-1} Y_{4e} \ast  |x|\right) = 1$, (\ref{slow}) follows. \qed

\section{Proof of Theorems \ref{theorem:leq1} and \ref{theorem:existence}}\label{sec:existence}

\indent As was shown in\-~(\ref{simpleq3}) and\-~(\ref{simpleq4}), there are at least two ways to write\-~(\ref{simpleq}) as a fixed point equation.
As it turns out, only the latter one
\begin{equation}
u(x) = \Phi(u)(x) := K_e (\mathcal V(x) +2e\rho u*u(x))
\label{iteration0}
\end{equation}
is adapted to solution by iteration, because of its monotonicity properties.
Starting with $u_0(x) = 0$, define
\begin{equation}
 u_n(x) = \Phi(u_{n-1})(x)
\end{equation}
for $n\geqslant1$. It is easy to see that for arbitrary $e,\rho \geqslant 0$, this produces a monotone increasing sequence of non-negative integrable functions. Thus, $u(x) := \lim_{n\to \infty}u_n(x)$ will exist, but it need not be integrable and it need not satisfy\-~(\ref{energy}) or\-~(\ref{con1}). 

\indent To bring\-~(\ref{energy}) into the iteration scheme, we take $e$ as the independent parameter, and define a sequence $\{\rho_n\}$ along with the sequence $\{u_n(x)\}$, both depending on $e$, through
\begin{equation}
 u_n(x)=K_e \mathcal V(x)+2e\rho_{n-1}K_e u_{n-1}\ast u_{n-1}(x) \ .
 \quad
 u_0(x)=0
 \label{iteration}
\end{equation}
and
\begin{equation}
 \rho_n:=\frac{2e}{\int(1-u_n(x))\mathcal V(x)} .
 \label{rhon}
\end{equation} 
Comparing\-~(\ref{iteration}) to\-~(\ref{iteration0}), note that the analog of $\Phi$ now depends on $n$.

\begin{Lemma}\label{lem1} Let $\mathcal{V} \in L^1(\mathbb{R}^d)\cap L^p(\mathbb{R}^d)$, $p>\max\{\frac d2,1\}$. Both sequences $\{\rho_n\}$ and $\{u_n\}$ are well defined and increasing, and for all $n$,
\begin{equation}\label{simple17}
\int_{\mathbb{R}^d} u_n dx < \frac{1}{2e}\int_{\mathbb{R}^d}\mathcal{V}(1-u_n)dx \ .
\end{equation}
\end{Lemma}
\bigskip

{\bf Proof:}
We proceed by induction. By definition, $u_0 =0$ and $\rho_0 = 2e\left(\int_{\mathbb{R}^d}\mathcal V(x)dx\right)^{-1}$. Also by definition
$u_1 = K_e\mathcal V \geqslant u_0$ and $\rho_1 = 2e\left( \int \mathcal V(1- K_e\mathcal{V})dx \right)^{-1}$.
 As noted in the discussion between\-~(\ref{con4}) and\-~(\ref{con4B}), 
\begin{equation}
2\int_{\mathbb{R}^d} u_1dx = \frac{1}{e}\int_{\mathbb{R}^d}\mathcal{V}(1-u_1)dx \leqslant \ \frac{1}{e}\int_{\mathbb{R}^d}\mathcal{V}dx.
\end{equation}
Since $t\mapsto t^{-1}$ is monotone decreasing on $(0,\infty)$, this shows that $\rho_1 > \rho_0$, and that\-~(\ref{simple17}) holds for $n=1$. 
\bigskip

\indent Now suppose that $u_{n} \geqslant u_{n-1} \geqslant 0$, $\rho_{n} \geqslant \rho_{n-1} \geqslant 0$, and $\int_{\mathbb{R}^d} u_{n} dx < \frac{1}{2e}\int_{\mathbb{R}^d}\mathcal V(1-u_{n})$, all of which we have just verified for $n=1$. Then
\begin{equation}
 u_{n+1} = K_e\mathcal V + 2e \rho_n K_e u_{n}*u_{n}(x) \geqslant K_e\mathcal V + 2e\rho_{n-1} K_e u_{n-1}*u_{n-1}(x) = u_n(x)\ ,
\end{equation}
and then 
\begin{equation}
\int_{\mathbb{R}^d} \mathcal V(1-u_{n+1}) dx < \int_{\mathbb{R}^d}\mathcal V(1-u_{n}) dx \ .
\end{equation}
Integrating both sides of $u_{n+1} = G_e \mathcal {V}(1- u_{n+1}) + 2e \rho_n G_e u_{n}*u_{n}$ yields,
\begin{equation}\label{simple15}
2\int_{\mathbb{R}^d} u_{n+1} dx = \frac{1}{2e}\int_{\mathbb{R}^d}\mathcal V(1-u_{n+1}) + \rho_n \left( \int_{\mathbb{R}^d}u_{n} dx \right)^2
\end{equation}
Then since
\begin{equation}
  int_{\mathbb{R}^d} u_{n} dx < \frac{1}{2e}\int_{\mathbb{R}^d}\mathcal V(1-u_{n}) = \frac{1}{\rho_n}
\end{equation}
(\ref{simple15}) implies 
\begin{equation}
 2\int_{\mathbb{R}^d} u_n dx \leqslant \frac{1}{2e}\int_{\mathbb{R}^d}\mathcal{V}(1-u_n) + \int_{\mathbb{R}^d}u_{n-1} dx\ .
\end{equation}
Then because $\int_{\mathbb{R}^d} u_{n} dx < \int_{\mathbb{R}^d} u_{n+1} dx $, we have
\begin{equation}
  \int_{\mathbb{R}^d} u_{n+1} dx < \frac{1}{2e}\int_{v}\mathcal{V}(1-u_{n+1}).
\end{equation}
This proves\-~(\ref{simple17}) for $n+1$, and shows that 
\begin{equation}
 0 \leqslant \frac{1}{2e}\int_{\mathbb{R}^d}\mathcal{V}(1-u_{n+1})dx \leqslant \frac{1}{2e}\int_{\mathbb{R}^d}\mathcal{V}(1-u_n)dx \ ,
\end{equation}
and then, as before, $\rho_{n+1}\geqslant \rho_n$. 
\qed
\bigskip

\begin{Lemma}\label{lem2} Let $\mathcal{V} \in L^1(\mathbb{R}^d)\cap L^p(\mathbb{R}^d)$, $p>\max\{\frac d2,1\}$. Then for all $n$ and $x$, $u_n(x)$ is continuous, vanishing at infinity, and $0 \leqslant u_n(x) \leqslant 1$. 
\end{Lemma}
\bigskip

{\bf Proof:} First consider $n=1$. Since $u_n = K_e\mathcal{V}$ with $\mathcal{V}\in L^p(\mathbb{R}^d)$, $u_1\in W^{2,p}(\mathbb{R}^d)$ and 
\begin{equation}
\Delta u_1(x) = \mathcal{V}(x)(u_1(x) - 1) + 4e u_1(x)\ .
\end{equation}
Since $K_e$ maps $L^p(\mathbb R^d)$ into $W^{2,p}(\mathbb R^d)$, $u_1$ is continuous and vanishes at infinity. Let $A := \{x\ :\ u_1(x) > 1\}$. Then $A$ is open. If $A$ is non-empty, then $u_1$ is subharmonic on $A$, and hence takes on its maximum on the boundary of $A$. Since $u_1$ would equal $1$ on the boundary, this is impossible, and $A$ is empty. This proves the assertion for $n=1$. 
\bigskip

\indent Now make the inductive hypothesis that $0 \leqslant u_n(x) \leqslant 1$ for all $x$. Then
\begin{equation}
  \|u_n\|_p^p \leqslant \|u_n\|_1 \leqslant\frac{1}{2e}\int_{\mathbb{R}^d}\mathcal{V}dx.
\end{equation}
By Young's inequality, $\|u_n\ast u_n\|_p \leqslant \|u_n\|_p\|u_1\|_1$, and hence 
$\mathcal{V} + 2e\rho_n u_n\ast u_n \in L^p(\mathbb{R}^d)$. Therefore, 
$u_{n+1}= K_e(\mathcal{V} + 2e\rho_n u_n\ast u_n) \in W^{2,p}(\mathbb{R}^d)$. It follows as before that $u_{n+1}$ is continuous and vanishing at infinity, and in particular, bounded, and 
\begin{eqnarray*}
\Delta u_{n+1}(x) &=& \mathcal{V}(x)(u_n(x) - 1) + 4e u_n(x) - 2e\rho_n u_n\ast u_n\\
&\geqslant& \mathcal{V}(x)(u_n(x) - 1) + 4e u_n(x) -2e\rho_n\|u_n\|_1 \|u_n\|_\infty\\
& \geqslant& 
 \mathcal{V}(x)(u_n(x) - 1) + 4e u_n(x) -2e
 \end{eqnarray*}
 where we have used $\rho_n\|u_n\|_1 \leqslant 1$, which is valid on account of\-~(\ref{simple17}). Define $A := \{x:\ u_{n+1}(x) > 1\}$. Then $u_{n+1}$ is subharmonic on $A$, and maximal on the boundary of $A$, where $u_n(x)$ would equal 
 $1$. This contradiction shows that $\|u_{n+1}\|_\infty \leqslant 1$.
\qed
\bigskip

\begin{Lemma}\label{lem3} Let $\mathcal{V} \in L^1(\mathbb{R}^d)\cap L^p(\mathbb{R}^d)$, $p>\max\{\frac d2,1\}$. 
Now let
\begin{equation}
 u(x): = \lim_{n\to\infty}u_n(x) \quad{\rm and}\quad \rho(e) = \lim_{n\to\infty}\rho_n(e)\ .
\end{equation}
Then both limits exist, $u\in W^{2,p}(\mathbb{R}^d)$ and $u$ satisfies\-~(\ref{simpleq}), (\ref{energy}) and\-~(\ref{con1}). 
\end{Lemma}
\bigskip

{\bf Proof:} 
By Lemma~\ref{lem1}, both limits exist, and by\-~(\ref{simple17}), $\rho(e) \leqslant \left(\int_{\mathbb{R}^d} K_e\mathcal{V}dx\right)^{-1}$.
Also by Lemma~\ref{lem1}, $\int_{\mathbb{R}^d} \leqslant \frac{1}{2e}\int_{\mathbb{R}^d}\mathcal{V}(x)dx$,
$u$ is integrable and $\lim_{n\to\infty}\|u_n - u\|_1 = 0$. Moreover, by Lemma~\ref{lem2}, $0 \leqslant u \leqslant 1$, and then $\|u\|_p^p \leqslant \|u\|_1$ and $\|u_n- u\|_p^p \leqslant (p+1)\|u_n-u\|_1$, and then by Young's Inequality
\begin{equation}
 \|u\ast u - u_n\ast u_n\|_p \leqslant \|u_n\|_1\|u_n -u\|_p + \leqslant \|u\|_1\|u_n -u\|_p\ .
\end{equation}
Therefore, $\lim_{n\to\infty}(\mathcal{V} +2e\rho_n(e) u_n\ast u_n) = (\mathcal{V} +2e\rho(e) u\ast u)$
with convergence in $L^p(\mathbb{R}^d)$. Then $\lim_{n\to\infty}K_e (\mathcal{V} +2e\rho_n(e) u_n\ast u_n) = K_e(\mathcal{V} +2e\rho(e) u\ast u)$ with convergence in $W^{2,p}(\mathbb{R}^d)$, and in particular, in
$L^p(\mathbb{R}^d)$. It now follows that $u = K_e(\mathcal{V} +2e\rho(e) u\ast u) $, and by the Dominated Convergence Theorem, the constraint $\rho = \frac{1}{2e}\int_{\mathbb{R}^d}\mathcal{V}(1-u)dx$ is satisfied.
By remarks made above, this means that $u$ satisfies\-~(\ref{simpleq}) and\-~(\ref{energy}). \qed
\bigskip

\begin{Lemma}\label{lem4}
 For all $e\in (0,\infty)$, the solution $u$ of the system\-~(\ref{simpleq}) and\-~(\ref{energy}) that we have constructed by iteration in Lemma~\ref{lem3} is the unique non-negative integrable solution for $\rho = \rho(e)$.
 Moreover, there does not exist such any such solution when $\rho \neq \rho(e)$. 
\end{Lemma}
\bigskip

{\bf Proof:} 
 Consider any non-negative solution integrable $\tilde u$, with
 \begin{equation}
 \tilde\rho=\frac{2e}{\int(1-\tilde u(x))\mathcal V(x)\ dx}
 .
 \end{equation}
 We first show that $\tilde u\geqslant u_n$ by induction.
 We have
 \begin{equation}
 \tilde u(x)-u_n(x)=2eK_e(\tilde\rho\tilde u\ast\tilde u(x)-\rho_{n-1}u_{n-1}\ast u_{n-1}(x))
 \end{equation}
 Since $u_0 =0$, the positivity of $\tilde u$ implies the positivity of $\tilde u(x) - u_1(x)$.
 If $\tilde u\geqslant u_{n-1}$, then, by\-~(\ref{rhon}), $\tilde\rho\geqslant\rho_{n-1}$, from which $\tilde u\geqslant u_n$ follows easily.
 This proves that both $\tilde\rho\geqslant\rho$ and $\tilde u\geqslant u$. However, integrating both sides of the latter inequality yields 
 \begin{equation}
 \frac{1}{\tilde \rho(e)} = \int\tilde u(x)\ dx \geqslant\int u(x)\ dx=\frac1{\rho(e)}\ .
 \end{equation}
 Since $\tilde\rho\geqslant\rho$, equality must hold, and then since $\tilde u\geqslant u$, it must be that 
 so $u=\tilde u$.
\qed
\bigskip 
 
\begin{Lemma}\label{lem5}
 The function $\rho(e)$ is continuous on $(0,\infty)$, with
 \begin{equation}
 \lim_{e\to 0}\rho(e) = 0
 ,\quad
 \lim_{e\to\infty}\rho(e) = \infty.
 \end{equation}
 In particular, for each $\rho\in (0,\infty)$, there is at least one $e\in (0,\infty)$ such that $\rho = \rho(e)$. 
\end{Lemma}
\bigskip

{\bf Proof:}
 We now turn to the continuity of $e\to \rho(e)$.
 For $n\in \mathbb{N}$, define functions $a_n(e)$ and $b_n(e)$ by
 \begin{equation}
 a_n:=\int u_n(x,e)\ dx \quad{\rm and}\quad b_n(e) = \frac1{2e}\int (1-u_n(x,e))\mathcal V(x)\ dx\ .
 \end{equation} 
 where we have temporarily made the dependence of $u_n$ on $e$ explicit. 
 Note that $b_n(e) = 1/\rho_n(e)$. 
 $u_1(x,e) = K_e\mathcal{V}$ is continuous in $e$ (and monotone decreasing) for each $x$. A simple induction shows that $u_n(x,e)$ is continuous in $e$ for each $x$.
 Then since $(1-u_n(x,e))\mathcal V(x) \leqslant \mathcal{V}(x)$, the Dominated Convergence Theorem yields the continuity of $\rho_n(e)$ for each $n$.
 Writing our iteration in the equivalent form (as in\-~(\ref{simpleq3})):
 \begin{equation}\label{simpleq3B}
u_n(x,e) = Y_{4e}*(\mathcal V (1- u_n(x,e))) +2e \frac{1}{b_{n-1}(e)} Y_{4e}\ast u_{n-1}\ast u_{n-1}(x,e) \ ,
\end{equation}
and integrating, we obtain
 \begin{equation}\label{simpleq3C} 
2a_n(x) = b_n(e) + \frac{1}{b_{n-1}(e)} a_{n-1}^2(e) \ ,
\end{equation}
Now an easy induction shows that $a_n(e)$ is continuous for each $n$. By\-~(\ref{simple17}), for each $n$,
 \begin{equation}
 a_n(e)\leqslant\frac{1}{\rho(e)}\leqslant b_n(e) \ .
 \end{equation}
 By Lemma\-~\ref{lem1}, as $n$ increases to infinity, $a_n(e)$ increases to $1/\rho(e)$, while $b_n(e) $ decreases to $1/\rho(e)$. It remains to show that this convergence is uniform on any compact interval in $(0,\infty)$.
 By\-~(\ref{simpleq3C}), 
 \begin{equation}\label{telescope}
 \frac{1}{b_n(e)}(a_n(e) - b_n(e))^2 = \frac{a_n^2(e)}{b_n(e)} -( 2a_n(e) - b_n(e)) = \frac{a_n^2(e)}{b_n(e)} - \frac{a_{n-1}^2(e)}{b_{n-1}(e)} \ .
 \end{equation}
 Sum both sides over $n\in \mathbb{N}$. The sum on the right telescopes, and since for all $e$, $a_0^2/b_0 = 0$ while $\lim_{n\to\infty}a_n^2(e)/b_n(e) = 1/\rho_n(e)$, 
 \begin{equation}
 \sum_{n=1}^\infty \frac{1}{b_n(e)}(a_n(e) - b_n(e))^2 = \frac{1}{\rho(e)} \ .
 \end{equation}
 By the bounds on $b_(e) = 1/\rho_n(e)$ and $\rho(e)$ provided by Lemma~\ref{lem1}, for all $e>0$,
 \begin{equation}
 \sum_{n=1}^\infty (a_n(e) - b_n(e))^2 \leqslant\frac{ \int \mathcal{V}dx}{\int K_e\mathcal{V}dx}\ ,
 \end{equation}
 and on any compact interval $[e_1,e_2]$, the right hand side is uniformly bounded by $C$, its value at $e_2$. Then since the summand on the left is monotone decreasing in $n$, we obtain for each $n$ that 
 \begin{equation}\label{rate}
 (a_n(e) - b_n(e))^2 \leqslant \frac{C}{n}
 \end{equation} uniformly on $[e_1,e_2]$. This proves the desired uniform convergence, and hence the continuity of $\rho(e)$. The final statement now follows from 
 (\ref{con4B}). \qed
\bigskip
 
{\bf Remark}: Note that $\|u - u_n\|_1 = \frac{1}{\rho} - a_n$, and hence by\-~(\ref{rate}), $\|u - u_n\|_1 \leqslant Cn^{-1/2}$. 
In fact, numerically, we find that the rate is significantly faster than this. For example, with $\mathcal V(x)=e^{-|x|}$ and $e=10^{-4}$, $\|u - u_n\|_1$ decays at least as fast as $n^{-3.5}$.
\bigskip

{\bf Proof of Theorem\-~\ref{theorem:leq1}} This theorem follows from Lemmas\-~\ref{lem2}, \ref{lem3} and\-~\ref{lem4}. \qed
\bigskip

{\bf Proof of Theorem\-~\ref{theorem:existence}} Every statement in the theorem has been established in Lemma~\ref{lem1} through Lemma~\ref{lem5}. \qed

\bigskip

We close this section by remarking that if $\mathcal{V}$ is radially symmetric, then so is $u_1 = K_e\mathcal{V}$, and then by a simple induction, so is $u_n$, hence also $u$, the unique solution $u$ provided by Theorem\-~\ref{theorem:existence}.
This is consistent with the first remark following Theorem\-~\ref{theorem:existence}.

\section{Asymptotics}\label{sec:asymptotics}
\indent In this section, we prove Theorem\-~\ref{theorem:asymptotics}.
We will first prove the high density asymptote\-~(\ref{largerho}), and then proceed to the low density\-~(\ref{3d}).
\bigskip

\indent
By Theorem\-~\ref{theorem:existence}, for each $\rho>0$ there exists at least one $e$ such that $\rho(e)=\rho$.
If there is more than one, the theorems proved in this section apply to every such solution.
Throughout this section, let $u_\rho$ denote the solution provided by Theorem\-~\ref{theorem:existence} and any such choice of $e$.
\bigskip

\subsection{High density $\rho$}

\begin{Lemma}[high density asymptotics]\label{lemma:large}
 If $\mathcal V$ is integrable, then as $\rho\to\infty$,
 \begin{equation}
 e=\frac\rho2\left(\int \mathcal V(x)\ dx\right)(1+o(1))
 .
 \end{equation}
\end{Lemma}
\bigskip

{\bf Remark}: From\-~(\ref{energy}),
\begin{equation}
 e\leqslant\frac\rho2\int\mathcal V(x)\ dx
 .
\end{equation}
Note that this is not an optimal bound, as follows from\-~(\ref{con4}).
\bigskip

{\bf Proof}:
 By\-~(\ref{energy}), it suffices to prove that
 \begin{equation}
 \lim_{\rho\to\infty}\int u_\rho(x)\mathcal V(x)\ dx
 =0
 .
 \label{limintuv}
 \end{equation}
 Let
 \begin{equation}
 \chi_\gamma:=\{x:\ \mathcal V(x)\geqslant \gamma\}
 \end{equation}
 and decompose
 \begin{equation}
 \int u_\rho(x)\mathcal V(x)\ dx
 =
 \int_{\chi_\gamma} u_\rho(x)\mathcal V(x)\ dx
 +
 \int_{\mathbb R^d\setminus\chi_\gamma} u_\rho(x)\mathcal V(x)\ dx
 \end{equation}
 which, by\-~(\ref{con2}), is bounded as follows
 \begin{equation}
 \int u_\rho(x)\mathcal V(x)\ dx
 \leqslant
 \int_{\chi_\gamma}\mathcal V(x)\ dx
 +
 \frac \gamma\rho
 .
 \end{equation}
 Since $\mathcal V$ is integrable, $\int_{\chi_\gamma}\mathcal V(x)\ dx\to0$ as $\gamma\to\infty$.
 Therefore,
 \begin{equation}
 \mathop{\mathrm{inf}}_{\gamma>0}
 \left(
 \int_{\chi_\gamma}\mathcal V(x)\ dx
 +
 \frac \gamma\rho
 \right)
 \mathop{\longrightarrow}_{\rho\to\infty}0
 .
 \end{equation}
\qed

\subsection{Low density $\rho$}\label{subsec:lowrho} In this section, we only consider the dimension $d=3$. As before, we suppose that $\mathcal{V}\in L^1(\mathbb{R}^3)\cap L^p(\mathbb{R}^3)$, $p> 3/2$, and $\mathcal{V} \geqslant 0$.

\indent We first recall the definition of the {\it scattering length} of the potential $\mathcal{V}$, and relate it to the solution of the system\-~(\ref{simpleq})-(\ref{energy}) The {\it scattering equation} is defined as
\begin{equation}
 -\Delta\varphi(x)=(1-\varphi(x))\mathcal V(x)
 ,\quad
 \lim_{|x|\to\infty}\varphi(x)=0\ .
 \label{scattering}
\end{equation}
Note that\-~(\ref{scattering}) can be written as $(-\Delta + \mathcal V)\varphi = \mathcal{V}$, and hence the solution is
\begin{equation}\label{scattering2}
 \varphi(x) = \lim_{e\downarrow 0} K_e \mathcal{V}(x) = \lim_{e\downarrow 0}u_1(x,e)\ ,
 \end{equation}
 where $u_1$ is the first term of the iteration introduced in the previous section. It follows from Lemma~\ref{lem2} that $0 \leqslant \varphi(x) \leqslant1$ for all $x$. 
 
\indent We now impose a mild localization hypothesis on $\mathcal{V}$: For 
$R>0$ define $\mathcal{V}_R(x) = \mathcal{V}(x)$ for $|x| > R$ and otherwise $\mathcal{V}_R(x) =0$. We require that for some $q>1$ and all sufficiently large $R$,
\begin{equation}\label{local}
\|\mathcal{V}_R\|_1 < R^{-q} \quad{\rm and}\quad \|\mathcal{V}_R \|_p < R^{-q} \ .
\end{equation}
By the lemma below, $\lim_{|x|\to\infty}|x|\varphi(x)$ exists. The {\em scattering length} $a$ is
 defined to be (in dimension $d=3$).
 \begin{equation}\label{sl}
 a= \lim_{|x|\to \infty} |x| \varphi(x).
 \end{equation}
For more information on the scattering length, see\-~\cite[appendix A]{LY01}.

\begin{Lemma}\label{sctlem} Let $\mathcal{V}\in L^1(\mathbb{R}^3)\cap L^p(\mathbb{R}^3)$, $p> 3/2$, and suppose that
the localization condition\-~(\ref{local}) is satisfied. Let $\varphi$ be the corresponding scattering solution given by\-~(\ref{scattering2}). Then the scattering length $a := \lim_{|x|\to\infty}\varphi(x)$ exists and satisfies
\begin{equation}\label{ephi}
4\pi a = \int \mathcal{V}(x)(1-\varphi(x))dx\ .
\end{equation}. 
\end{Lemma} 

{\bf Proof:} By the resolvent identity, 
$ \varphi(x) = G\ast (\mathcal{V}(1- \varphi))(x)$
where $G(x) = \frac{1}{4\pi|x|}$. Since $p> 3/2$. $p' < 3$, and it is easy to decompose $G$ into the sum of two pieces,
$G = G_1+G_2$ where $G_1 \in L^{p'}(\mathbb{R}^d)$ and $G_2 \in L^{4}(\mathbb{R}^d)$.
Then for all $R$ sufficiently large,
\begin{equation}
 0 \leqslant G\ast (\mathcal{V_R}(1- \varphi))(x) \leqslant (\|G_1\|_{p'} + \|G_2\|_4) R^{-q}\ .
\end{equation}
For $0 < r < 1$, then for $|y| < r|x|$, 
${\displaystyle \frac{1}{1+r} \leqslant \frac{|x|}{|x-y|} \leqslant \frac{1}{1-r}}$.
It follows that for all sufficiently large $|x|$, 
\begin{equation}
\frac{1}{1+r}\int_{|y|< r|x|}\mathcal{V}(y)(1-\varphi(y)) dy+ o(1) \leqslant 4\pi|x|\varphi(x) \leqslant 
 \frac{1}{1-r}\int_{|y|< r|x|}\mathcal{V}(y)(1-\varphi(y)) dx + o(1)\ .
\end{equation}
Taking $|x|\to \infty$, and then $r \to 0$ proves\-~(\ref{local}).\qed
\bigskip

{\bf Remark}: The following lemma is valid if the scattering length $a$ were {\em defined} by (\ref{ephi}).
For this reason, we do not impose the additional  condition (\ref{local}) in the statement of Theorem\-~\ref{theorem:asymptotics}: Lemma\-~\ref{sctlem} reconciles the stated definition with the formula (\ref{ephi}).
\bigskip

\begin{Lemma}[low density asymptotics]\label{lemma:small}
 If $\mathcal V$ is non-negative and integrable and $d=3$, then
 \begin{equation}
 e=2\pi\rho a\left(1+\frac{128}{15\sqrt\pi}\sqrt{\rho a^3}+o(\sqrt\rho)\right)
 .
 \label{3dlemma}
 \end{equation}
\end{Lemma}
\bigskip

{\bf Proof}:
 The scheme of the proof is as follows.
 We first approximate the solution $u$ by $w$, which is defined as the decaying solution of
 \begin{equation}
 -\Delta w_\rho(x)=(1-u_\rho(x))\mathcal V(x)
 .
 \label{u1}
 \end{equation}
 The energy of $w_\rho$ is defined to be
 \begin{equation}
 e_w:=\frac\rho2\int(1-w_\rho(x))\mathcal V(x)\ dx
 \label{ew}
 \end{equation}
 and, as we will show, it is {\it close} to $e$, more precisely,
 \begin{equation}
 e-e_w=\frac{16\sqrt 2 e^{\frac32}}{15\pi^2}\int\mathcal V(x)\ dx+o(\rho^{\frac32})
 .
 \label{eew}
 \end{equation}
 In addition, (\ref{u1}) is quite similar to the scattering equation\-~(\ref{scattering}).
 In fact we will show that $e_w$ is {\it close} to the energy $2\pi\rho a$ of the scattering equation
 \begin{equation}
 e_w-2\pi\rho a
 =
 -\frac{16\sqrt 2e^{\frac32}}{15\pi^2}\int \varphi(x) \mathcal V(x)\ dx+o(\rho^{\frac32})
 .
 \label{ewephi}
 \end{equation}
 Summing\-~(\ref{eew}) and\-~(\ref{ewephi}), we find
 \begin{equation}
 e=2\pi\rho a\left(1+\frac{32\sqrt 2 e^{\frac32}}{15\pi^2\rho}+o(\sqrt\rho)\right)
 ,
 \end{equation}
 from which\-~(\ref{3dlemma}) follows.
 We are thus left with proving\-~(\ref{eew}) and\-~(\ref{ewephi}).
 \bigskip

 \point{\bf Proof of\-~(\ref{eew})}.
 By\-~(\ref{energy}) and\-~(\ref{ew}),
 \begin{equation}
 e-e_w=\frac\rho2\int(w_\rho(x)-u_\rho(x))\mathcal V(x)\ dx
 .
 \end{equation}
 We will work in Fourier space
 \begin{equation}
 \hat u_\rho(k):=\int e^{ikx}u_\rho(x)\ dx
 \label{fourieru}
 \end{equation}
 which satisfies, by\-~(\ref{simpleq}),
 \begin{equation}
 (k^2+4e)\hat u_\rho(k)
 =\frac{2e}\rho S(k)
 +2e\rho \hat u^2(k)
 \end{equation}
 with
 \begin{equation}
 S(k):=\frac\rho{2e}\int e^{ikx}(1-u_\rho(x))\mathcal V(x)\ dx
 .
 \label{S}
 \end{equation}
 Therefore,
 \begin{equation}
 \hat u_\rho(k)=\frac1\rho\left(\frac{k^2}{4e}+1-\sqrt{\left(\frac{k^2}{4e}+1\right)^2-S(k)}\right)
 .
 \label{hatu}
 \end{equation}
 Similarly, the Fourier transform of $w_\rho$ is
 \begin{equation}
 \hat w_\rho(k):=\int e^{ikx}w_\rho(x)\ dx=\frac{2e S(k)}{\rho k^2}
 .
 \label{u1hat}
 \end{equation}
 Note that, as $|k|\to\infty$, $\hat u\sim \frac{2eS(k)}{\rho k^2}$, so, while $\hat u_\rho$ is not integrable, $\hat u_\rho-\hat w_\rho$ is.
 We invert the Fourier transform:
 \begin{equation}
 u_\rho(x)-w_\rho(x)=\frac1{8\pi^3\rho}\int e^{-ikx}
 \left(\frac{k^2}{4e}+1-\sqrt{\left(\frac{k^2}{4e}+1\right)^2-S(k)}-\frac{2eS(k)}{k^2}\right)\ dk
 .
 \end{equation}
 We change variables to $\tilde k:=\frac k{2\sqrt e}$:
 \begin{equation}
 u_\rho(x)-w_\rho(x)
 =
 \frac{e^{\frac32}}{\rho\pi^3}\int e^{-i2\sqrt e\tilde kx}
 \left(\tilde k^2+1-\sqrt{(\tilde k^2+1)^2-S(2\tilde k\sqrt e)}-\frac{S(2\tilde k\sqrt e)}{2\tilde k^2}\right)\ d\tilde k
 .
 \end{equation}
 Furthermore,
 \begin{equation}
 s\mapsto\left|\tilde k^2+1-\sqrt{(\tilde k^2+1)^2-s}-\frac{s}{2\tilde k^2}\right|
 \end{equation}
 is monotone increasing.
 In addition, by~\-(\ref{S}) and~\-(\ref{simpleq}), and using the fact that $u_\rho(x)\leqslant 1$ (see Lemma\-~\ref{lem2}) and $\mathcal V(x)\geqslant 0$,
 \begin{equation}
 |S(k)|\leqslant\frac\rho{2e}\int|(1-u_\rho(x))\mathcal V(x)|\ dx=1
 .
 \end{equation}
 Therefore
 \begin{equation}
 \left|
 \tilde k^2+1-\sqrt{(\tilde k^2+1)^2-S(2\tilde k\sqrt e)}-\frac{S(2\tilde k\sqrt e)}{2\tilde k^2}
 \right|
 \leqslant
 \left|
 \tilde k^2+1-\sqrt{(\tilde k^2+1)^2-1}-\frac{1}{2\tilde k^2}
 \right|
 .
 \end{equation}
 Therefore,
 \begin{equation}
 |u_\rho(x)-w_\rho(x)|\leqslant
 \frac{e^{\frac32}}{\rho\pi^3}\int\left|\tilde k^2+1-\sqrt{(\tilde k^2+1)^2-1}-\frac{1}{2\tilde k^2}\right|\ d\tilde k
 =\frac{32\sqrt 2 e^{\frac32}}{15\pi^2\rho}
 .
 \end{equation}
 By dominated convergence, and using the fact that $S(0)=1$,
 \begin{equation}
 \begin{largearray}
 \lim_{e\to 0}\frac1{e^{\frac32}}(e-e_w)
 =-\lim_{e\to 0}\frac\rho{2e^{\frac32}}\int (u_\rho(x)-w_\rho(x))\mathcal V(x)\ dx
 \\[0.5cm]\hfill
 =
 -\frac12\int\mathcal V(x)\left(
	\frac1{\pi^3}\int\left(\tilde k^2+1-\sqrt{(\tilde k^2+1)^2-1}-\frac1{2\tilde k^2}\right)d \tilde k
 \right)\ dx
 =\frac{16\sqrt 2}{15\pi^2}\int\mathcal V(x)\ dx
 .
 \end{largearray}
 \end{equation}
 Using\-~(\ref{con4C}), this proves\-~(\ref{eew}).
 Incidentally, again by dominated convergence,
 \begin{equation}
 u_\rho(x)-w_\rho(x)=
 \frac{e^{\frac32}}{\rho\pi^3}\int\left(\tilde k^2+1-\sqrt{(\tilde k^2+1)^2-1}-\frac{1}{2\tilde k^2}\right)\ d\tilde k
 =-\frac{32\sqrt 2 e^{\frac32}}{15\pi^2\rho}+\sqrt\rho f_\rho(x)
 \label{uapproxw}
 \end{equation}
 with
 \begin{equation}
 0\leqslant f_\rho(x)\leqslant\frac{32\sqrt2 e^{\frac32}}{15\pi^2\rho}
 ,\quad
 f_\rho(x)\mathop{\longrightarrow}_{\rho\to0}0
 \end{equation}
 pointwise in $x$.

 \bigskip

 \point{\bf Proof of\-~(\ref{ewephi})}.
 Let
 \begin{equation}
 \xi(r):=w_\rho(r)-\varphi(r).
 \label{xi}
 \end{equation}
 By\-~(\ref{u1}), (\ref{scattering}) and\-~(\ref{simpleq}),
 \begin{equation}
 (-\Delta+\mathcal V(x))\xi(x)=-(u_\rho(x)-w_\rho(x))\mathcal V(x)
 .
 \end{equation}
 Therefore, by\-~(\ref{ephi}),
 \begin{equation}
 e_w-2\pi\rho a=-\frac\rho2\int \xi(x)\mathcal V(x)\ dx
 =-\frac\rho2\int\mathcal V(x)(-\Delta+\mathcal V)^{-1}((u-w)\mathcal V)(x)\ dx
 \end{equation}
 and
 \begin{equation}
 (-\Delta+\mathcal V)^{-1}\mathcal V(x)=\varphi(x)
 \end{equation}
 so
 \begin{equation}
 e_w-2\pi\rho a
 =-\frac\rho2\int\varphi(x)(u_\rho(x)-w_\rho(x))\mathcal V(x)\ dx
 .
 \end{equation}
 By\-~(\ref{uapproxw}),
 \begin{equation}
 e_w-2\pi\rho a
 =\frac{16\sqrt 2e^{\frac32}}{15\pi^2}\int\varphi(x)\mathcal V(x)\ dx
 -\frac{\rho^{\frac32}}2\int\varphi(x)f_\rho(x)\mathcal V(x)\ dx
 .
 \end{equation}
 Since $x\mapsto f_\rho(x)$ is bounded, we can use dominated convergence to show\-~(\ref{ewephi}).
\qed

\section{Decay of $u$}\label{sec:decay}
\indent In this section, we prove Theorem\-~\ref{theorem:decay}.
Our proof assumes that $\mathcal V$ decays exponentially, because we will use analyticity properties of the Fourier transform of the potential $\mathcal V$.
In particular, the theorem holds if $\mathcal V$ has compact support.
We expect the result to hold for any potential that decays faster than $|x|^{-4}$.
Algebraic decay for $u$ seems natural: by\-~(\ref{simpleq}), $u\ast u$ must decay at infinity in the same way as $u$.
This is the case if $u$ decays algebraically, but would not be so if, say, it decayed exponentially.
\bigskip

{\bf Proof of theorem\-~\ref{theorem:decay}}:
We begin by proving (\ref{gendecay}) in arbitrary dimension. Recall that the first part has already been proved in Theorem~\ref{positivity} without the additional assumption on the potential. For the second part, recall that by the first remark after Theorem~\ref{theorem:existence},  $u$ is also radial, and hence $\mathcal{V}(1-u)$ is non-negative and radial. It then follows from the hypotheses on $\mathcal{V}$ that $g :=2\rho eY_{4e}\ast Y_{4e}*[\mathcal{V}(1-u)]$
satisfies
\begin{equation}
  \int |x|^2 g(x) dx < \infty \quad{\rm and}\quad \int x g(x) d x = 0\ .
\end{equation}
Then, as explained in Section~\ref{sec:pos}, if $f := 2e\rho Y_{4e}\ast u$, $f - f\ast f = g\geqslant 0$, and then by \cite[Theorem 4]{CJLL20}, the second part of  (\ref{gendecay}) follows.   Note that if 
\begin{equation}
   u(|x|)\mathop\sim_{|x|\to\infty}\frac\alpha{|x|^m}
\end{equation}
for some $\alpha>0$, then the only choice of $m$ that is consistent with (\ref{gendecay}) is $m = d+1$.   
\medskip

We now specialize to $d=3$, and impose the additional assumption on the potential.
\medskip

Recall that the Fourier transform of $u$ (\ref{fourieru}) satisfies\-~(\ref{hatu}):
  \begin{equation}
    \hat u(|k|)=\frac1\rho\left(\frac{k^2}{4e}+1-\sqrt{\left(\frac{k^2}{4e}+1\right)^2-S(|k|)}\right)
  \end{equation}
  where $S$ was defined in\-~(\ref{S}):
  \begin{equation}
    S(|k|):=\frac\rho{2e}\int e^{ikx}(1-u(|x|))\mathcal V(|x|)\ dx
    .
  \end{equation}
  We split
  \begin{equation}
    \hat u(|k|)=\widehat{\mathcal U}_1(|k|)+\widehat{\mathcal U}_2(|k|)
    \label{hatusplit}
  \end{equation}
  with
  \begin{equation}
    \widehat{\mathcal U}_1(|k|):=\frac{2eS(|k|)}{\rho(1+k^2)}
  \end{equation}
  so that, taking the large $|k|$ limit in\-~(\ref{hatu}),
  \begin{equation}
    \widehat{\mathcal U}_2(|k|)=O(|k|^{-4}S^2(|k|))
    \label{apriori_U2}
  \end{equation}
  so $\widehat{\mathcal U}_2$ is integrable.
  \bigskip

  \point{\bf Decay of $\mathcal U_1$}. We first show that
  \begin{equation}
    \mathcal U_1(|x|):=\frac1{(2\pi)^3}\int e^{-ikx}\widehat{\mathcal U}_1(|k|)\ dk
  \end{equation}
  decays exponentially in $|x|$. We have
  \begin{equation}
    \mathcal U_1(|x|)=(-\Delta+1)^{-1}(1-u(|x|))\mathcal V(|x|)
    =Y_1\ast((1-u)\mathcal V)(|x|)
  \end{equation}
  with
  \begin{equation}
    Y_1(|x|):=\frac{e^{-|x|}}{4\pi|x|}
    .
  \end{equation}
  Therefore, by\-~(\ref{expdecay}),
  \begin{equation}
    \mathcal U_1(|x|)
    \leqslant
    \frac A{4\pi}\int_{|y|>R} \frac{e^{-|x-y|-B|y|}}{|x-y|}\ dy
    +
    \frac 1{4\pi}\int_{|y|<R} \frac{e^{-|x-y|}}{|x-y|}\mathcal V(|y|)\ dy
  \end{equation}
  so, denoting $b:=\min(B,1)$,
  \begin{equation}
    \mathcal U_1(|x|)\leqslant
    \frac A{4\pi}\int \frac{e^{-b(|x-y|+|y|)}}{|x-y|}\ dy
    +
    \frac{e^{-(|x|-R)}}{4\pi(|x|-R)}\int\mathcal V(|y|)\ dy
  \end{equation}
  and since
  \begin{equation}
    \frac A{4\pi}\int \frac{e^{-b(|x-y|+|y|)}}{|x-y|}\ dy
    =
    \frac{Ae^{-b|x|}}{4b^2}(b|x|+1)
  \end{equation}
  we have
  \begin{equation}
    \mathcal U_1(|x|)\leqslant
    \frac{Ae^{-b|x|}}{4b^2}(b|x|+1)
    +
    \frac{e^{-(|x|-R)}}{4\pi(|x|-R)}\int\mathcal V(|y|)\ dy
    .
    \label{U1}
  \end{equation}
  \bigskip

  \point{\bf Analyticity of $\mathcal U_2$}.
  We now turn to
  \begin{equation}
    \mathcal U_2(|x|):=\frac1{(2\pi)^3}\int e^{-ikx}\widehat{\mathcal U}_2(|k|)\ dk
    =\frac1{4i\pi^2|x|}\sum_{\eta=\pm}\eta\int_0^\infty e^{i\eta\kappa|x|}\kappa\widehat{\mathcal U}_2(\kappa)\ d\kappa
    .
    \label{U2}
  \end{equation}
  We start by proving some analytic properties of $\widehat{\mathcal U}_2$, which, we recall from\-~(\ref{hatu}) and\-~(\ref{hatusplit}), is
  \begin{equation}
    \widehat{\mathcal U}_2(|k|)=\frac1\rho\left(\frac{k^2}{4e}+1-\sqrt{\left(\frac{k^2}{4e}+1\right)^2-S(|k|)}-\frac{2eS(|k|)}{1+k^2}\right)
    .
  \end{equation}
  \bigskip

  \subpoint First of all, $S$ is analytic in a strip about the real axis:
  \begin{equation}
    S(\kappa)=4\pi\int_0^\infty \mathrm{sinc}(\kappa r)r^2\mathcal V(r)(1-u(r))\ dr
    ,\quad
    \mathrm{sinc}(\xi):=\frac{\sin(\xi)}\xi
  \end{equation}
  so
  \begin{equation}
    \partial^nS(\kappa)=4\pi\int_0^\infty \partial^n\mathrm{sinc}(\kappa r)r^{n+2}\mathcal V(r)(1-u(r))\ dr
    .
  \end{equation}
  We will show that if $\mathcal Im(\kappa)\leqslant \frac B2$ (the factor $\frac12$ can be improved to any factor that is $<1$, but this does not matter here), then there exists $C>0$ which only depends on $A$ and $B$ such that
  \begin{equation}
    |\partial^nS(\kappa)|\leqslant n!C^n
    .
    \label{decaydS}
  \end{equation}
  As a consequence, $S$ is analytic in a strip around the real line of height $\frac B2$.
  In particular, if we define the strip
  \begin{equation}
    H_\tau:=\{z:\ |\mathcal Im(z)|\leqslant r^{-\tau},\ \mathcal Re(z)>0\}
  \end{equation}
  with $0<\tau<1$, and take
  \begin{equation}
    r>\left(\frac B2\right)^{-\frac1\tau}
  \end{equation}
  then $S$ is analytic in $H_\tau$.
  \bigskip

  \bigskip

  \subsubpoint We now prove\-~(\ref{decaydS}).
  We first treat the case $|\kappa|\leqslant\frac B2$. We have
  \begin{equation}
    \mathrm{sinc}(\xi)=\sum_{p=0}^\infty\frac{(-1)^p\xi^{2p}}{(2p+1)!}
  \end{equation}
  so
  \begin{equation}
    \partial^n\mathrm{sinc}(\xi)=\sum_{p=\lceil\frac n2\rceil}^\infty\frac{(-1)^p\xi^{2p-n}}{(2p+1)(2p-n)!}
    .
  \end{equation}
  Therefore
  \begin{equation}
    |\partial^n\mathrm{sinc}(\xi)|\leqslant
    \sum_{p=\lceil\frac n2\rceil}^\infty\frac{|\xi|^{2p-n}}{(2p-n)!}
    \leqslant \cosh(|\xi|)
    .
  \end{equation}
  Thus,
  \begin{equation}
    |\partial^n S(\kappa)|\leqslant 4\pi\int_0^\infty \cosh(|\kappa|r)r^{n+2}\mathcal V(r)(1-u(r))\ dr
  \end{equation}
  so, by\-~(\ref{expdecay}),
  \begin{equation}
    |\partial^n S(\kappa)|\leqslant
    4A\pi\int_R^\infty \cosh(|\kappa|r)r^{n+2}e^{-Br}\ dr
    +
    4\pi\int_0^R \cosh(|\kappa|r)r^{n+2}\mathcal V(r)\ dr
  \end{equation}
  and
  \begin{equation}
    |\partial^n S(\kappa)|\leqslant
    8A\pi\int_0^\infty r^{n+2}e^{-(B-|\kappa|)r}\ dr
    +
    8\pi e^{|\kappa|R}R^{n}\int r^2\mathcal V(r)\ dr
  \end{equation}
  which, if $|\kappa|\leqslant \frac B2$, implies that
  \begin{equation}
    8A\pi\int_0^\infty r^{n+2}e^{-(B-|\kappa|)r}\ dr
    \leqslant
    8A\pi\int_0^\infty r^{n+2}e^{-\frac B2r}\ dr
    =
    \frac{2^{n+6}A\pi}{B^{n+3}}(n+2)!
  \end{equation}
  and
  \begin{equation}
    8\pi e^{|\kappa|R}R^{n+2}\int\mathcal V(r)\ dr
    \leqslant
    8\pi e^{\frac B2R}R^{n}\int r^2\mathcal V(r)\ dr
  \end{equation}
  which implies\-~(\ref{decaydS}) in this case.
  \bigskip

  \subsubpoint We now turn to $|\kappa|\geqslant \frac B2$:
  \begin{equation}
    \partial^n\mathrm{sinc}(\xi)=\sum_{p=0}^n{n\choose p}\partial^p\sin(\xi)\frac{(n-p)!(-1)^{n-p}}{\xi^{n-p+1}}
  \end{equation}
  so
  \begin{equation}
    |\partial^n\mathrm{sinc}(\xi)|\leqslant 2e^{\mathcal Im(\xi)}\sum_{p=0}^n\frac{n!}{p!}|\xi|^{-(n-p+1)}
    .
  \end{equation}
  Therefore,
  \begin{equation}
    |\partial^nS(\kappa)|\leqslant
    8\pi \sum_{p=0}^n\frac{n!}{p!|\kappa|^{n-p+1}}
    \int_0^\infty e^{\mathcal Im(\kappa)r}r^{p+1}\mathcal V(r)(1-u(r))\ dr
  \end{equation}
  so, by\-~(\ref{expdecay}),
  \begin{equation}
    |\partial^nS(\kappa)|\leqslant
    \sigma_1+\sigma_2
  \end{equation}
  with
  \begin{equation}
    \sigma_1:=8A\pi\sum_{p=0}^n\frac{n!}{p!|\kappa|^{n-p+1}}
    \int_R^\infty r^{p+1}e^{-(B-\mathcal Im(\kappa))r}\ dr
  \end{equation}
  and
  \begin{equation}
    \sigma_2:=8\pi\sum_{p=0}^n\frac{n!}{p!|\kappa|^{n-p+1}}
    \int_0^R r^{p+1}e^{\mathcal Im(\kappa)r}\mathcal V(r)\ dr
    .
  \end{equation}
  Furthermore,
  \begin{equation}
    \sigma_1
    =
    8A\pi n!
    \sum_{p=0}^n\frac{p+1}{(B-\mathcal Im(\kappa))^{p+2}|\kappa|^{n-p+1}}
  \end{equation}
  so, as long as $|\kappa|\geqslant \frac12B$ and $\mathcal Im(\kappa)\leqslant\frac12B$,
  \begin{equation}
    \sigma_1
    \leqslant
    \frac{2^{n+6}A\pi}{B^{n+3}}n!\sum_{p=0}^n(p+1)
    =
    \frac{2^{n+5}A\pi}{B^{n+3}}(n+2)!
    .
  \end{equation}
  In addition,
  \begin{equation}
    \sigma_2
    \leqslant
    8\pi\sum_{p=0}^n\frac{n!}{p!|\kappa|^{n-p+1}}R^{p-1}e^{\mathcal Im(\kappa)R}\int_0^R r^2\mathcal V(r)\ dr
  \end{equation}
  so
  \begin{equation}
    \sigma_2
    \leqslant
    8\pi\sum_{p=0}^n\frac{n!2^{n-p+1}}{p!B^{n-p+1}}R^{p-1}e^{\mathcal Im(\kappa)R}\int_0^R r^2\mathcal V(r)\ dr
    \leqslant
    \frac{2^{n+4}\pi}{RB^{n+1}} n!e^{RB}\int_0^R r^2\mathcal V(r)\ dr
  \end{equation}
  which implies\-~(\ref{decaydS}) in this case.
  \bigskip

  \subpoint We have thus proved that $S$ is analytic in $H_\tau$, which implies that the singularities of $\widehat{\mathcal U}_2$ in $H_\tau$ all come from the branch points of $\sqrt{F(|k|)}$ with $F(|k|):=(\frac{k^2}{4e}+1)^2-S(|k|)$.
  For $\kappa\in\mathbb R$,
  \begin{equation}
    |S(\kappa)|\leqslant 1
  \end{equation}
  so, for $\kappa\in\mathbb R$,
  \begin{equation}
    F(\kappa)\geqslant \frac{\kappa^2}{2e}
    .
  \end{equation}
  Therefore, since $F$ is analytic in a strip around the real axis, there exists an open set containing the real axis in which $F$ has one and only one root, at $0$.
  Thus the only branch point of $\sqrt F$ on the real axis is $0$.
  Thus, $\widehat{\mathcal U}_2$ is analytic in $H_\tau$.
  \bigskip

  \point{\bf Decay of $\mathcal U_2$}. We deform the integral to the path
  \begin{equation}
    \{i\eta y,\ 0<y<|x|^{-\tau}\}\cup\{i\eta |x|^{-\tau}+y,\ y>0\}
  \end{equation}
  and find
  \begin{equation}
    \int_0^\infty e^{i\eta\kappa|x|}\kappa\widehat{\mathcal U}_2(\kappa)\ d\kappa
    =
    I_1+I_2
    \label{I12}
  \end{equation}
  with
  \begin{equation}
    I_1:=
    -\int_0^{|x|^{-\tau}}e^{-y|x|}y\widehat{\mathcal U}_2(i\eta y)\ dy
  \end{equation}
  and
  \begin{equation}
    I_2:=
    e^{-|x|^{1-\tau}}
    \int_0^\infty e^{i\eta y|x|}(i\eta|x|^{-\tau}+y)\widehat{\mathcal U}_2(i\eta|x|^{-\tau}+y)\ dy
    .
  \end{equation}
  \bigskip

  \subpoint We first estimate $I_1$.
  We expand $S$:
  \begin{equation}
    S(\kappa)=1-\beta \kappa^2+O(|\kappa|^4)
  \end{equation}
  with $\beta>0$ (since $S$ is analytic and symmetric, and $|S(|k|)|\leqslant 1$).
  Therefore, $y\mapsto\widehat{\mathcal U}_2(iy)$ is $\mathcal C^2$ for $y\neq0$, and
  \begin{equation}
    \widehat{\mathcal U}_2(i\eta y)
    =
    \frac1\rho-\frac{i\eta y}\rho\sqrt{\frac1{2e}+\beta}
    +O(y^2)
    .
  \end{equation}
  Furthermore,
  \begin{equation}
    -\int_0^{|x|^{-\tau}}e^{-y|x|}y\ dy
    =
    -\frac1{|x|^2}+\frac{1+|x|^{1-\tau}}{|x|^2}e^{-|x|^{1-\tau}}
  \end{equation}
  \begin{equation}
    -\int_0^{|x|^{-\tau}}e^{-y|x|}y^2\ dy
    =
    -\frac2{|x|^3}+\frac{1+|x|^{1-\tau}(2+x^{1-\tau})}{|x|^3}e^{-|x|^{1-\tau}}
  \end{equation}
  and
  \begin{equation}
    -\int_0^{|x|^{-\tau}}e^{-y|x|}y^3\ dy
    =
    O(|x|^{-4})
  \end{equation}
  \begin{equation}
    I_1=
    -\frac1{\rho|x|^2}
    +\frac{2i\eta}{\rho|x|^3}\sqrt{\frac1{2e}+\beta}
    +O(|x|^{-4})
  \end{equation}
  so
  \begin{equation}
    \frac1{4i\pi^2|x|}\sum_{\eta=\pm}\eta I_1=\frac1{\pi^2\rho|x|^4}\sqrt{\frac1{2e}+\beta}
    +O(|x|^{-5})
    .
    \label{ineqI1}
  \end{equation}
  \bigskip

  \subpoint We now bound $I_2$.
  Recall that, for $\kappa\in\mathbb R$, $|S(\kappa)|\leqslant 1$.
  Recalling\-~(\ref{decaydS}),
  \begin{equation}
    |S(\kappa+i\eta|x|^{-\tau})|
    \leqslant
    \sum_{n=0}^\infty\frac1{n!}|\partial^nS(\kappa)|^n|x|^{-n\tau}
    \leqslant
    \frac1{1-C|x|^{-\tau}}
    \leqslant 2
  \end{equation}
  provided $|x|^\tau>2C$.
  Therefore, for large $\kappa$, by\-~(\ref{apriori_U2}),
  \begin{equation}
    |\widehat{\mathcal U}_2(\kappa+i\eta)|=O(\kappa^{-4})
  \end{equation}
  so
  \begin{equation}
    I_2\leqslant C'e^{-|x|^{1-\tau}}
    \label{ineqI2}
  \end{equation}
  for some constant $C'>0$.
  \bigskip

  \subpoint Inserting\-~(\ref{ineqI1}) and\-~(\ref{ineqI2}) into\-~(\ref{I12}) and\-~(\ref{U2}), we find that
  \begin{equation}
    \mathcal U_2(|x|)=\frac1{\pi^2\rho|x|^4}\sqrt{\frac1{2e}+\beta}+O(|x|^{-5})
  \end{equation}
  which, using\-~(\ref{U1}), concludes the proof of the theorem.
  \bigskip
\qed

\section{Comparison with the Bose gas}\label{sec:bosegas}
\subsection{Sketch of the derivation of the simple equation}
\indent The simple equation\-~(\ref{simpleq})-(\ref{energy}) was originally derived\-~\cite{Li63} to approximate the ground state energy $E_0$ of a repulsive Bose gas, which is a system of $N$ quantum particles interacting via the repulsive potential $\mathcal V$.
The ground state energy of this system is the lowest eigenvalue of the Hamiltonian operator
\begin{equation}
 H_N:=-\frac12\sum_{i=1}^N\Delta_i+\sum_{1\leqslant i<j\leqslant N}\mathcal V(x_i-x_j)
\end{equation}
acting on the space of $L_2$ functions on the torus $\mathbb T_V$ of volume $V$.
The corresponding eigenfunction, which we will denote by $\psi_N$, satisfies
\begin{equation}
 H_N\psi_N(x_1,\cdots,x_N)=E_0\psi_N(x_1,\cdots,x_N)
 \label{eigenvalue}
\end{equation}
with $x_i\in\mathbb T_V$.
As is well known, by a Perron-Frobenius argument, $\psi_N$ is unique, non-negative, and hence symmetric under exchanges $x_i\leftrightarrow x_j$, and under translations.
\bigskip

\indent We can write $E_0$ by integrating both sides of\-~(\ref{eigenvalue}):
\begin{equation}
 E_0=\frac {N(N-1)}{2V}\int g_N^{(2)}(x)\mathcal V(x)\ dx
 \label{energyg}
\end{equation}
with
\begin{equation}
 g_N^{(p)}(x_1,\cdots,x_p):=\frac{V^j}{\int\psi_N(x_1,\cdots,x_N)\ dx_1\cdots dx_N}\int\psi_N(x_1,\cdots,x_N)\ dx_{p+1}\cdots dx_N
 \label{g}
\end{equation}
and $g_N^{(2)}(x_1,x_2)\equiv g_N^{(2)}(x_1-x_2)$.
The computation of $E_0$ thus reduces to that of $g_N^{(2)}$.
Note that the kinetic energy does not appear explicitly in\-~(\ref{energyg}).
\bigskip

\indent To compute $g_N^{(2)}$, integrate both sides of\-~(\ref{eigenvalue}) with respect to $x_3,\cdots,x_N$.
This yields an equation relating $g_N^{(2)}$, $g_N^{(3)}$ and $g_N^{(4)}$.
The main approximation made in\-~\cite{Li63}, is to write $g_N^{(3)}$ and $g_N^{(4)}$ products of $g_N^{(2)}$ factors: roughly,
\begin{equation}
 g_N^{(p)}(x_1,\cdots,x_p)\approx\prod_{1\leqslant i<j\leqslant p}g_N^{(2)}(x_i-x_j)
 .
 \label{approxprod}
\end{equation}
This is a sensible approximation in the case of low density $\rho=\frac NV\ll 1$.
Indeed, in this regime, one might expect $\psi_N$ to be approximately a {\it Bijl-Dingle-Jastrow function}:
\begin{equation}
 \psi_N(x_1,\cdots,x_N)\approx\prod_{1\leqslant i<j\leqslant N}e^{-\phi(x_i-x_j)}
\end{equation}
for some appropriately chosen real function $\phi$.
Thus, $\psi_N$ is approximated by the {\it partition function} of a classical statistical mechanical model of particles interacting via the pair-potential $\phi$.
In this setting, $g_N^{(p)}$ is the {\it $p$-point correlation function} of the canonical Gibbs distribution of this model.
When\-~(\ref{approxprod}) holds asymptotically as the particles move away from each other (remember, the density is low), the statistical mechanics system is said to satisfy the {\it clustering property}.
There is a long literature on proving the clustering property for a large class of potentials $\phi$, see, among many others, \cite{Ru99,Ga99,PT12}.
\bigskip

\indent Assuming the clustering property for the potential $\phi$, the assumption\-~(\ref{approxprod}) does not seem far fetched.
This product structure leads to an equation for $g_N^{(2)}$.
At this stage, one takes the thermodynamic limit: $N\to\infty$ and $\rho=\frac NV$ fixed.
There are some subtleties to taking this limit, which are explained in\-~\cite{Li63}.
Defining $u:=1-g_\infty^{(2)}$, the equation for $u$ is\-~\cite[(3.29)]{Li63}.
After a few extra reasonable approximations, this equation reduces to\-~(\ref{simpleq}).
The equation for the energy\-~(\ref{energy}) is simply the $N\to\infty$ limit of\-~(\ref{energyg}).
\bigskip

\indent In particular, $u$ is related to the correlation function $g^{(2)}$ of the Bose gas.
The condition\-~(\ref{con1}) that $u(x)\leqslant1$ is necessary to ensure that $g^{(2)}(x)\geqslant0$.
However, $u(x)\geqslant0$ is not a physical requirement, as $g^{(2)}(x)$ could, in principle, be $>1$
for some $x$.

\subsection{Numerical comparison}\label{subsec:numerics}
\indent One of the motivations for studying the simple equation is that it provides a simple tool to approximate the ground state energy of the Bose gas.
In\-~\cite{LL64}, it was found that in one dimension the simple equation gives a value for the energy that differs from the Bose gas ground state energy by at most 69\% (a more complete form of the equation yields an even better result, with a maximal error of 19\%).
In one dimension, the difference is larger at high density.
\bigskip

\begin{figure}
 \hfil\includegraphics[width=10cm]{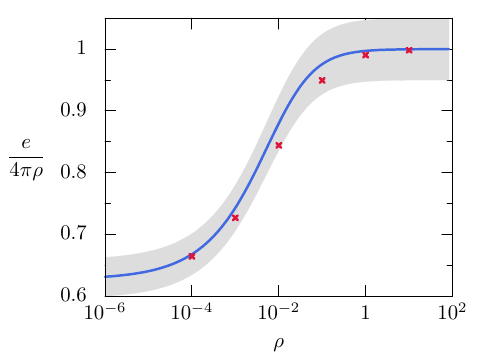}
 \caption{
 Plot of $\frac e{4\pi\rho}$ as a function of $\rho$ on a log scale.
 The potential is $\mathcal V(r)=e^{-r}$, in which case the scattering length is $a\approx 1.25$.
 The solid curve is the energy computed from the simple equation\-~(\ref{simpleq})-(\ref{energy}), and the discrete points are the values of the energy of the Bose gas computed by M.\-~Holzmann\-~\cite{CHe} using a Monte Carlo algorithm.
 The gray area corresponds to a 5\% error on the value of the energy.
 At low densities, we recover the Lenz asymptote $\frac e{4\pi\rho}\sim\frac a2$ and at high densities, we recover $\frac e{4\pi\rho}\sim1$.
 The difference between the Monte Carlo simulation and the solution of the simple equation is smaller than 5\%.
 }
 \label{fig:bosegas}
\end{figure}

\indent In three dimensions, by Theorem\-~\ref{theorem:asymptotics}, the simple equation predicts the correct low density asymptote as the Bose gas.
This is a not so surprising, since the derivation of the simple equation from the ground state equation of the Bose gas sketched above seems somewhat sensible when the density is low.
However, when the density is high, at least in the case in which the potential has a non-negative Fourier transform, the simple equation also yields the same asymptote as the Bose gas.
In fact, considering the case
\begin{equation}
 \mathcal V(x)=e^{-|x|}
\end{equation}
(which has a positive Fourier transform), we compared the ground state energy of the simple equation with values from a Monte Carlo simulation of the Bose gas computed by M.\-~Holzmann\-~\cite{CHe}, to whom we are most grateful for sharing his unpublished work.
The comparison is in figure\-~\ref{fig:bosegas}, in which we found that the maximal error made by the simple equation, over the {\it entire range of densities}, is 5\%!
This is a promising result, which we will investigate in more depth and with more rigor in a later publication.
\bigskip

\section{Open problems and conjectures}\label{sec:open}

\point{\bf Monotonicity}.
An important open problem is to show that $e\mapsto\rho(e)$ is an increasing function.
If the solution of the simple equation is in any way related to the ground state wave function of the Bose gas, then this should hold: if the density increases, the energy should increase.
In addition, it would enable us to prove the uniqueness of the solution of the simple equation with fixed $\rho$, and might even allow us to generalize our result to potentials with hard core components, as well as to relax the constraint that $\mathcal V$ decays exponentially in Theorem\-~\ref{theorem:decay}.
By running a few numerical computations, it seems clear that $\rho(e)$ should be increasing, see figure\-~\ref{fig:bosegas}.
Using a modified iteration in which $\rho$ is fixed, we have proved that $e\rho(e)$ is strictly monotone increasing in $e$, but the proof that $\rho(e)$ is as well has eluded us thus far.
\bigskip

\begin{figure}
 \hfil\includegraphics[width=10cm]{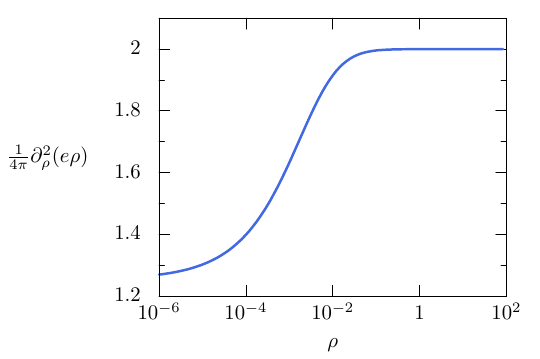}
 \caption{
 Numerical evaluation of $\frac1{4\pi}\partial_\rho^2(\rho e)$ for $\mathcal V(r)=e^{-r}$.
 The asymptotic values are, for $\rho\to0$, $a\approx 1.25$ and for $\rho\to\infty$, $2$.
 This second derivative seems to be clearly positive, so $\rho e$ appears to be convex.
 }
 \label{fig:convexity}
\end{figure}

\point{\bf Convexity}.
Another open problem is to prove that $\rho e(\rho)$ {\it is a convex function}, or, equivalently, that $\frac1{\rho(e)}$ is convex.
In a physical setting, one expects $\rho e(\rho)$ to be convex.
Indeed if $\rho e=:e_{\mathrm v}$ were {\it not convex}, there would exist $\rho_1<\rho<\rho_2$ such that $\frac{\rho_1+\rho_2}2=\rho$ and $e_{\mathrm v}(\rho_1)+e_{\mathrm v}(\rho_2)<2e_{\mathrm v}(\rho)$.
Furthermore, $e_{\mathrm v}$ is the energy per unit volume, and, considering a volume $V$ that is split into two equal halves, we find that a configuration in which one half of the volume holds a density $\rho_1$ of particles, whereas the other holds $\rho_2$ would have energy
\begin{equation}
 \frac V2(e_{\mathrm v}(\rho_1)+e_{\mathrm v}(\rho_2))<Ve_{\mathrm v}(\rho)
 .
\end{equation}
Therefore, it would pay to have more particles in one half than in the other, which is unstable.
Numerically, it seems quite clear that $\rho e(\rho)$ is convex, see figure\-~\ref{fig:convexity}.
\bigskip

\point {\bf Solution of the full equation}.
The simple equation\-~(\ref{simpleq}) is actually a simplified version of an equation that should approximate the Bose gas more accurately\-~\cite{Li63}:
\begin{equation}
  (-\Delta+\mathcal V(x))u(x)
  =\mathcal V(x)
  -\rho(1-u(x))(2K(x)-\rho L(x))
  \label{fulleq}
\end{equation}
with
\begin{equation}
  K(x):=u\ast S(x)
  ,\quad
  S(x):=(1-u(x))\mathcal V(x)
\end{equation}
\begin{equation}
  L(x):=
  \int u(y)u(z-x)\left(
    1-u(z)-u(y-x)+\frac12u(z)u(y-x)
    \ dydz
  \right)S(y)
  .
  \label{L}
\end{equation}
Note that $e$ appears only as the integral of $S$, see\-~(\ref{energy}).
While little is known rigorously about this equation, we have been studying it numerically in collaboration with M. Holzmann \cite{CHe}, and have found it to be remarkably accurate.
These results will be detailed in a future publication.

\vskip20pt

{\bf Acknowledgements}:
  {\it
  Thanks go to Markus Holzmann for his hard work and skill computing the ground state energy of the Bose gas, and for sharing his results with us.
  E.C. gratefully acknowledges support through NSF grant DMS-174625.
  I.J. gratefully acknowledges support through NSF grant DMS-1802170.
  }
\bigskip

%\bibliographystyle{abbrv}
%\bibliography{bibliography}

\end{document}